\pdfoutput=1
\RequirePackage{ifpdf}
\ifpdf 
\documentclass[pdftex]{sigma}
\else
\documentclass{sigma}
\fi

\def\half{{1\over 2}}

\numberwithin{equation}{section}

\begin{document}

\newcommand{\arXivNumber}{1411.6170}

\allowdisplaybreaks

\renewcommand{\thefootnote}{$\star$}

\renewcommand{\PaperNumber}{029}

\FirstPageHeading

\ShortArticleName{Exact M-Theory Solutions, Integrable Systems, and Superalgebras}

\ArticleName{Exact M-Theory Solutions, Integrable Systems,\\
and Superalgebras\footnote{This paper is a~contribution to
the Special Issue on Exact Solvability and Symmetry Avatars in honour of Luc Vinet.
The full collection is available at
\href{http://www.emis.de/journals/SIGMA/ESSA2014.html}{http://www.emis.de/journals/SIGMA/ESSA2014.html}}}

\Author{Eric D'HOKER}

\AuthorNameForHeading{E.~D'Hoker}

\Address{Department of Physics and Astronomy, University of California, Los Angeles, CA 90095, USA}
\Email{\href{mailto:dhoker@physics.ucla.edu}{dhoker@physics.ucla.edu}}
\URLaddress{\url{http://www.pa.ucla.edu/directory/eric-dhoker}}

\ArticleDates{Received January 05, 2015, in f\/inal form April 03, 2015; Published online April 11, 2015}

\Abstract{In this paper, an overview is presented of the recent construction of fully back-reacted half-BPS solutions in
11-dimensional supergravity which correspond to near-horizon geometries of M2 branes ending on, or intersecting with, M5
and M5$'$ branes along a~self-dual string.
These solutions have space-time manifold ${\rm AdS}_3 \times S^3 \times S^3$ warped over a~Riemann surface~$\Sigma$, and
are invariant under the exceptional Lie superalgebra $D(2,1;\gamma) \oplus D(2,1;\gamma)$, where $\gamma $ is a~real
continuous parameter and $|\gamma|$ is governed by the ratio of the number of M5 and M5$'$ branes.
The construction proceeds by mapping the reduced BPS equations
onto an integrable f\/ield theory on~$\Sigma$ which is of the Liouville sine-Gordon type.
Families of regular solutions are distinguished by the sign of~$\gamma$, and include a~two-parameter Janus solution for
$\gamma >0$, and self-dual strings on M5 as well as asymptotically ${\rm AdS}_4/{\mathbb Z}_2$ solutions for $\gamma
<0$.}


\Classification{81Q60; 17B80}

\renewcommand{\thefootnote}{\arabic{footnote}}
\setcounter{footnote}{0}

\section{Introduction}

The main theme of my collaboration with Luc Vinet in the mid 1980s was the study of dynamical supersymmetries and
associated Lie superalgebras in certain integrable quantum mechanical systems involving magnetic monopoles and dyons.
In our f\/irst joint paper~\cite{D'Hoker:1983xu}, we showed that the standard Pauli equation for a~non-relativistic spinor
in the presence of a~background Dirac magnetic monopole exhibits a~dynamical supersymmetry.
The corresponding supercharges close onto the conformal symmetry of the Dirac magnetic monopole, thereby producing the
Lie superalgebra ${\rm OSp}(1|2)$.
In a~series of subsequent papers~\cite{D'Hoker:1985kb, D'Hoker:1985et}, we extended these results to integrable systems
which include a~dyon as well as~$1/r^2$ and~$1/r$ potentials, established the presence of associated dynamical
supersymmetries and higher rank Lie superalgebras, and we solved the spectra using purely group theoretic methods.

The main theme of the present paper is related to the subject of my earlier work with Luc Vinet, in the sense that it
deals with integrable systems, dynamical supersymmetries, and Lie superalgebras, albeit now in the context of
11-dimensional supergravity instead of mechanical systems with a~f\/inite number of degrees of freedom.
Specif\/ically, we shall present an overview of recent work in which exact solutions with ${\rm SO}(2,2) \times {\rm
SO}(4) \times {\rm SO}(4)$ isometry and 16 residual supersymmetries (so-called half-BPS solutions) to 11-dimensional
supergravity are constructed on space-time manifolds of the form
\begin{gather}
\label{1a1}
\left({\rm AdS}_3 \times S^3 \times S^3\right)\ltimes \Sigma.
\end{gather}
The radii of the anti-de-Sitter space ${\rm AdS}_3$ and of the spheres $S^3$ are functions of the two-dimensional
Riemann surface~$\Sigma$, so that the product $\ltimes$ with~$\Sigma$ is warped.
The construction of~\cite{D'Hoker:2008wc} proceeds by reducing the BPS equations of 11-dimensional supergravity to the
space-time~\eqref{1a1}, and then mapping the reduced BPS equations onto
an integrable 2-dimensional f\/ield theory which is a~close cousin of the Liouville and sine-Gordon theories.
Although large families of exact solutions to this f\/ield theory, and correspondingly to the supergravity problem, have
been constructed in~\cite{D'Hoker:2008qm,janus,Estes:2012vm}, a~full understanding and analysis of its integrability
properties remains to be achieved.

The motivation for this work derives from M-theory, string theory, and gauge/gravity duality as realized by holography.
For introductions to these topics and their interrelation, we refer the reader
to~\cite{Aharony:1999ti,D'Hoker:2002aw,Duff,DZF}.
The basic constituents of M-theory are the M2 and M5 branes with world volumes of respective dimensions $2+1$ and $5+1$.
The metric and other f\/ields for one M2 brane, or for a~stack of parallel M2 branes, is known analytically in the
supergravity approximation.
The same holds for M5 branes.
But an analytical solution for the intersection of M2 and M5 branes continues to elude us, although solutions with
smeared branes have been obtained, and their form is surprisingly
simple~\cite{Gauntlett:1996pb,Tseytlin:1996bh,Youm:1999zs} (see also~\cite{Gauntlett:1998kc}).

In the work reviewed here, it is shown that half-BPS solutions of M2 branes which end on M5 branes or intersect with
them, may be constructed analytically and explicitly, in the near-horizon limit.

In this presentation to an audience of specialists in integrable systems and Lie algebras and superalgebras, however, we
shall emphasize integrability and group theory, and further expand upon those topics.
For the physical signif\/icance of the solutions, within the context of brane intersections and holography, we refer the
reader to the recent paper~\cite{Bachas:2013vza}, where these properties are addressed in full.

\section{M-theory synopsis}

M-theory unif\/ies the f\/ive critical superstring theories, namely Type I, IIA, IIB, Heterotic $E_8 \times E_8$, and
Heterotic ${\rm SO}(32)$, and provides a~natural geometric framework for the unif\/ication of the dualities between those
superstring theories and some of their compactf\/ications~\cite{Witten:1995ex}.
As M-theory contains gravity, its basic length scale~$\ell$ is set by Newton's constant.
Since M-theory provides a~complete unif\/ication, it has no dimensionless free couplings.

\looseness=1
M-theory permits a~perturbative treatment in a~space-time which itself has a~f\/inite typical length scale, which we shall designate by~$R$.
(Note that this condition is not fulf\/illed by 11-dimensional f\/lat Minkowski space-time.) The perturbation expansion
corresponds to the limit where~$R$ is either small or large compared to~$\ell$.
It is in those limits that M-theory is best-understood.
Two classic cases are as follows.
First, when 11-dimensional space-time is compactif\/ied on a~circle of radius~$R$, then M-theory coincides with Type IIA
superstring theory with string coupling $R/\ell$, and admits a~perturbative expansion for $R/\ell \ll 1$.
Second, when the f\/luctuations of the metric and other f\/ields are of a~typical length~$R$ which is large compared
to~$\ell$, M-theory admits a~perturbative expansion in powers of $\ell/R$ around 11-dimensional supergravity.

The Type IIA approximation to M-theory gives access to the dynamics of the theory at all energy scales, including very
high energy at the Planck scale, and is therefore of great conceptual value.
Most questions of physical importance, including compactif\/ication to 4 space-time dimensions, however, involve much
lower energy scales, for which the supergravity limit may be trusted.
Moreover, the 11-dimensional supergravity approximation to M-theory and its compactif\/ications faithfully preserves
dualities, and provides a~calculable framework for gauge/gravity duality via holography.

\subsection{M2 and M5 branes}

The fundamental constituents of M-theory are M2 branes and M5 branes.
An Mp brane (for $p=2,5$) is an extended object in M-theory whose worldvolume has dimension $1+p$, so that~$p$ is the
spacial dimension of the brane, while the additional 1 accounts for the time dimension on the brane.
In 11-dimensional supergravity, whose bosonic f\/ield contents consists of a~metric~$ds^2$ and a~4-form f\/ield strength
$F=dC$ as we shall see more explicitly below, M5 branes carry magnetic charge $N_5$ of~$F$, while M2 branes carry
electric charge, def\/ined by
\begin{gather}
\label{2a1}
N_2={1 \over 4 \pi^4} \oint_{{\cal C}_2}\left(\star F+\half C \wedge F\right), \qquad N_5={1 \over 2 \pi^2}
\oint_{{\cal C}_5} F
\end{gather}
for basic homology cycles ${\cal C}_2$ and ${\cal C}_5$ of dimensions 7 and 4 respectively.
These charges are quantized, so that $N_2$ and $N_5$ are integers.
Therefore, it makes sense to refer to the conf\/iguration with $N_2=1$ as a~single M2 brane and to $N_2 >1 $ as a~stack of
$N_2$ parallel M2 branes.
Analogously, $N_5=1$ is the single M5 brane, while $N_5>1$ corresponds to s stack of $N_5$ parallel M5 branes.

Mp branes living in f\/lat Minkowski space-time are represented by fairly simple classical solutions which bear some
similarity to the Schwarzschild solution in pure gravity.
The Mp brane solutions have the geometry ${\mathbb R}^{1,p} \ltimes {\mathbb R}^{10-p}$, and are invariant under
Poincar\'e transformation on ${\mathbb R}^{1,p}$.
Corresponding to this product structure, we shall introduce coordinates $x^\mu$ along the brane with $\mu =0,1, \dots,
p$ as well as coordinates orthogonal to the brane which we represent by a~$10-p$ dimensional vector ${\bf y}$.
M2 and M5 branes respectively have the following metrics
\begin{gather}
{\rm M2}    \qquad   ds^2 =\left(1+{c_2 N_2 \ell^6 \over y^6}\right)^{-{2 \over 3}} dx^\mu dx_\mu +\left(1+
{c_2 N_2 \ell^6 \over y^6}\right)^{+{1 \over 3}} d{\bf y}^2,
\nonumber
\\
{\rm M5}    \qquad   ds^2 =\left(1+{c_5 N_5 \ell^3 \over y^3}\right)^{-{1 \over 3}} dx^\mu dx_\mu +\left(1+
{c_5 N_5 \ell^3 \over y^3}\right)^{+{2 \over 3}} d{\bf y}^2.
\label{2a2}
\end{gather}
Here, $y=|{\bf y}|$ is the f\/lat Euclidean distance to the brane; $dx^\mu$ is contracted with the f\/lat Minkowski metric
with signature $(-+\dots +)$ along the brane; and $c_2$, $c_5$ stand for constants which are independent of $N_2$ and~$N_5$.

Flat Minkowski space-time is a~solution to 11-dimensional supergravity with 32 Poincar\'e supercharges.
The M2 brane, or more generally, a~stack of parallel M2 branes, preserves 16 of those 32 Poincar\'e supersymmetries.
The same is true for a~stack of parallel M5 branes.

\subsection{Near-horizon geometry}

The M2 and M5 brane solutions in supergravity are regular, despite the apparent singularity of the metric at $y=0$.
To see this, we consider the near-horizon approximation $y^6 \ll c_2 N_2\ell^6 $ for M2 branes, and $y^3 \ll c_5 N_5
\ell^3 $ for M5 branes, in which the metrics reduce to the following expressions (after changing coordinates to $z \sim
\ell^3/y^2$ for M2 and $z^2\sim \ell^3 /y$ for M5)
\begin{gather}
{\rm M2}
\qquad
ds^2=(c_2 N_2)^{1 \over 3} \ell^2\left({1 \over 4} {dz^2+dx^\mu dx_\mu \over z^2}+{d{\bf y}^2-dy^2 \over y^2}\right),
\nonumber
\\
{\rm M5}
\qquad
ds^2=(c_5 N_5)^{2 \over 3} \ell^2\left(4 {dz^2+dx^\mu dx_\mu \over z^2}+{d{\bf y}^2-dy^2 \over y^2}\right).
\label{2b1}
\end{gather}
We recognize the metric of ${\rm AdS}_4 \times S^7$ with radii proportional to $(c_2N_2)^{1 \over 6} \ell$ for M2, and
the metric of ${\rm AdS}_7 \times S^4$ with radii proportional to $(c_5 N_5)^{1 \over 3} \ell$ for M5.
We remind the reader that these maximally symmetric spaces are the coset spaces $S^{d+1}={\rm SO}(d+2)/{\rm SO}(d+1)$
and ${\rm AdS}_{d+1}={\rm SO}(d,2)/{\rm SO}(d,1)$ for Minkowski signature $AdS$.
Note that when $N_2, N_5 \gg 1$, the radii of these spaces are large and the curvature is small in units of~$\ell$, so
that the supergravity approximation to M-theory is indeed trustworthy.
Finally, the cycles used to def\/ine the electric charge of the M2 brane and the magnetic charge of the M5 brane
in~\eqref{2a1} are homeomorphic to the spheres of the near-horizon limits of these branes, namely respectively ${\cal
C}_2=S^7$ and ${\cal C}_5=S^4$.

Gauge/gravity duality is the conjectured holographic equivalence of M-theory on the space-time ${\rm AdS}_4 \times S^7$
with a~3-dimensional conformal quantum f\/ield theory with 32 supercharges but without gravity.
This theory is known as ABJM theory, and admits a~standard Lagrangian
formulation~\cite{Aharony:2008ug, Bagger:2006sk,Gustavsson:2007vu}.
Analogously, a~6-dimensional conformal quantum f\/ield theory with 32 supercharges is expected to exist which is
holographically dual to M-theory on ${\rm AdS}_7 \times S^4$, but the nature of this theory is still unclear, and it is
unlikely that it admits a~(standard) Lagrangian formulation.

\section{Geometry and symmetries of intersecting branes} 

A stack of $N_2$ parallel M2 branes has 16 residual Poincar\'e supersymmetries, which in the near-horizon limit is
enhanced to 32 supersymmetries forming the Lie superalgebra ${\rm OSp}(8|4,{\mathbb R})$.
Analogously, a~stack of $N_5$ parallel M5 branes has 16 residual Poincar\'e supersymmetries, which in the near-horizon
limit become enhanced to 32 supersymmetries forming ${\rm OSp}(8^*|4)$.

\subsection{Intersecting branes with residual supersymmetry}

When space-time is populated with a~generic collection of M2 and M5 branes, the geometry of the branes will be altered
by gravitational and other forces of M-theory, and the population will generically preserve no residual supersymmetry.
For special angles between the branes, however, some degree of residual supersymmetry may be preserved.
We refer the reader to~\cite{Berman:2007bv,Boonstra:1998yu,Gauntlett:1997cv,Smith:2002wn,Tseytlin:1997cs} for helpful
overviews and references to earlier work.

The simplest example is, of course, when the branes are parallel, as we had already discussed earlier.
Another example is when the branes have certain mutually orthogonal directions, along with other parallel directions.
For a~collection of M2 and M5 branes, the simplest such example is obtained when the M2 and M5 branes have 2 parallel
directions, all others being orthogonal.
We may choose a~coordinate system in 11-dimensional space-time in which the M2 brane is along the 012 directions, and
the M5 brane along the 013456 directions.
This conf\/iguration is schematically represented in Table~\ref{table1}, where directions parallel to a~brane are indicated with
the letter~$x$, and the 10-th dimension of space is designated by $\natural=10$.

\begin{table}[htdp]\centering\caption{Half-BPS intersecting M2 and M5 brane conf\/iguration.}\label{table1}\vspace{1mm}
\begin{tabular}{|c||c|c|c|c|c|c|c|c|c|c|c|}
\hline
branes & 0 & 1 & 2 & 3 & 4 & 5 & 6 & 7 & 8 & 9 & $\natural$
\\
\hline
\hline
M2 & $x$ & $x$ & $x$ &&&&&&&&
\\
\hline
M5 & $x$ & $x$ & & $x$ & $x$ & $x$ & $x$ &&&&
\\
\hline
\end{tabular}
\end{table}

It may be shown that the conf\/iguration of Table~\ref{table1}, for a~collection of arbitrary numbers~$N_2$ and~$N_5$ of M2 and
M5-branes respectively, preserves 8~residual Poincar\'e supersymmetries, and is thus half-BPS.
Actually, the conf\/iguration of Table~\ref{table1} is not the most general half-BPS conf\/iguration of M2 and M5 branes.
Indeed, one may add a~stack of $N_5'$ parallel M5 branes in the direction $01789\natural$, which we shall denote by
M5$'$ as shown in Table~\ref{table2} below.
The full conf\/iguration with stacks of $N_2$, $N_5$, $N_5'$ M2, M5, and M5$'$ branes respectively preser\-ves~8 Poincar\'e
supersymmetries, and is the most general such BPS conf\/iguration.

\begin{table}[htdp]\centering \caption{General half-BPS intersecting M2 and M5 brane conf\/iguration.}\label{table2}
\vspace{1mm}
\begin{tabular}{|c||c|c|c|c|c|c|c|c|c|c|c|}
\hline
branes & 0 & 1 & 2 & 3 & 4 & 5 & 6 & 7 & 8 & 9 & $\natural$
\\
\hline
\hline
M2 & $x$ & $x$ & $x$ &&&&&&&&
\\
\hline
M5 & $x$ & $x$ & & $x$ & $x$ & $x$ & $x$ &&&&
\\
\hline
M5$'$ & $x$ & $x$ & &&&&& $x$ & $x$ & $x$ & $x$
\\
\hline
\end{tabular}
\end{table}

When $N_2$, $N_5$, $N_5' \gg 1$, one expects a~corresponding supergravity solution to exist, but no exact solution has been
obtained so far.
(Solutions are available when one or both of the stacks of branes are ``smeared''.) Our goal will be to obtain
supergravity solutions not for the entire system of M2 and M5 branes, but only for their near-horizon limit.

\subsection{Symmetries of M2 and M5 and their near-horizon geometry}

\looseness=1
The bosonic symmetries of the M2 and M5 branes separately, and of their half-BPS intersection may essentially be read
of\/f from Table~\ref{table2}.
The supersymmetric completion of these symmetries is further dictated by the requirement of 32 supercharges for M2 and
M5 separately, and 16 supercharges for their half-BPS intersection.
We shall begin here by discussing the symmetries of the M2 and M5 branes separately, leaving the case of intersections
to the subsequent subsection.

A single M2 brane, or a~stack of M2 branes, has a~Poincar\'e symmetry algebra ${\rm ISO}(2,1)$ along the M2 brane in the
$012$ directions, and ${\rm SO}(8)$ symmetry in the directions $3456789\natural$ orthogonal to the brane, giving in
total ${\rm ISO}(2,1) \oplus {\rm SO}(8)$.
In the near-horizon limit, the bosonic symmetry ${\rm SO}(8)$ is unchanged, but the Poincar\'e symmetry gets enhanced to
the conformal symmetry algebra in $2+1$ dimensions.
Using the isomorphism ${\rm SO}(2,3)={\rm Sp}(4,{\mathbb R})$, the full bosonic symmetry is then ${\rm SO}(8) \oplus
{\rm Sp}(4,{\mathbb R})$.
There is only one Lie superalgebra with this maximal bosonic subalgebra and 32 supercharges, namely ${\rm
OSp}(8|4,{\mathbb R})$, and it is the full Lie superalgebra symmetry of the near-horizon space-time ${\rm AdS}_4 \times
S^7$.

Similarly, a~single M5 brane or a~stack of M5 branes, has a~Poincar\'e symmetry ${\rm ISO}(1,5)$ along the directions
$013456$ of the M5 brane.
Using the isomorphism ${\rm SO}(5)={\rm Sp}(4)$, the symmetry in the directions $2789\natural$ is given by ${\rm
ISO}(1,5) \oplus {\rm Sp}(4)$.
In the near-horizon limit, this symmetry gets enhanced to ${\rm SO}(2,6) \oplus {\rm Sp}(4)$, and extends uniquely to
the Lie superalgebra ${\rm OSp}(2,6|4)={\rm OSp}(8^*|4)$, which is the symmetry of the near-horizon space-time ${\rm
AdS}_7 \times S^4$.

\subsection{Symmetries of half-BPS intersecting branes}

The symmetry algebras for the half-BPS intersection of M2 branes with M5 branes, or for the half-BPS intersection of M2
branes with M5$'$ branes, or for the half-BPS intersection of M2, M5, and M5$'$ branes are all the same, as may again be
derived by inspecting Table~\ref{table2}.
It is given by the Poincar\'e algebra ${\rm ISO}(1,1)$ along the branes in the $01$ directions, along with a~f\/irst ${\rm
SO}(4)$ in the directions $3456$ and a~second ${\rm SO}(4)$ in the directions $789\natural$, giving the total bosonic
symmetry ${\rm ISO}(1,1) \oplus {\rm SO}(4) \oplus {\rm SO}(4)$.
In the near-horizon limit, this symmetry gets enhanced to ${\rm SO}(2,2) \oplus {\rm SO}(4) \oplus {\rm SO}(4)$.

Which Lie superalgebras have 16 supercharges and ${\rm SO}(2,2) \oplus {\rm SO}(4) \oplus {\rm SO}(4)$ as maximal
bosonic subgroup? No simple Lie superalgebra qualif\/ies.
We might have anticipated this result by inspecting the part of space-time on which the bosonic algebra acts, which is
${\rm AdS}_3 \times S^3 \times S^3$.
This space is isomorphic to the Lie group $B={\rm SO}(2,1) \times {\rm SO}(3) \times {\rm SO}(3)$ (up to factors of
${\mathbb Z}_2$) with Lie algebra ${\cal B}= {\rm SO}(2,1) \oplus {\rm SO}(3) \oplus {\rm SO}(3)$.
The isometry algebra of the Lie group~$B$ is given by the commuting left and right actions of ${\cal B}$ on~$B$, as it
would be for any Lie group, thereby giving the isometry algebra ${\cal B} \oplus {\cal B}$.
The Lie superalgebra we are seeking should follow the same pattern, and should therefore be of the form ${\cal G} \oplus
{\cal G}$, with ${\cal B}$ the maximal bosonic subalgebra of~${\cal G}$.
Recasting ${\cal B}$ equivalently as ${\cal B}={\rm SO}(4) \oplus {\rm Sp}(2,{\mathbb R})$, it is manifest that a~f\/irst
candidate for~${\cal G}$ is ${\cal G}={\rm OSp}(4|2,{\mathbb R})$.
However, ${\cal B}$ may also be recast as ${\cal B}={\rm SO}(4^*) \oplus {\rm USp}(2)$, so an alternative candidate for
${\cal G}$ is given by ${\cal G}={\rm OSp}(4^*|2)$.

The above candidates are special cases of the general ${\cal G}$ which consists of the exceptional Lie superalgebra
${\cal G}=D(2,1;\gamma)$, specif\/ically its real form whose maximal bosonic subalgebra is ${\rm SO}(2,1) \oplus {\rm SO}(4)$.
The parameter $\gamma $ is real and non-zero.
In view of the ref\/lection property $D(2,1;\gamma^{-1})=D(2,1;\gamma)$ for this real form, the range of~$\gamma$ may be
restricted to the interval
\begin{gather*}
\gamma \in [-1,1].
\end{gather*}
This hypothesis f\/its nicely with the results of~\cite{Sevrin:1988ew}, where $D(2,1;\gamma)$ arose as one member in the
classif\/ication of possible 2-dimensional superconformal f\/ield theory invariance algebras.
For $\gamma=1$, the exceptional Lie superalgebra $D(2,1;\gamma)$ reduces to the classical Lie superalgebra ${\rm
OSp}(4|2,{\mathbb R})$, while for $\gamma=-1/2$ it reduces to ${\rm OSp}(4^*|2)$, so that we recover the earlier two
candidates.
To summarize, the Lie superalgebra which leaves the half-BPS intersection of M2 branes with M5 and M5$'$ branes invariant
is given~by
\begin{gather}
\label{3a2}
D(2,1;\gamma) \oplus D(2,1;\gamma).
\end{gather}
From the explicit supergravity solutions, to be discussed next, we will conf\/irm the symmetry under~\eqref{3a2}, and link
the parameter $|\gamma|$ to the ratio of the number of M5 and M5$'$ branes.

\section{BPS solutions in 11-dimensional supergravity}

Having developed the geometry and articulated symmetries of intersecting brane conf\/igurations, and of their near-horizon
limit, in the preceding section, we shall now move onto deriving exact solutions within the context of 11-dimensional
supergravity for these brane intersections, in the near-horizon limit.

\subsection{11-dimensional supergravity}

Supergravity in 11-dimensional space-time has 32 supersymmetries and has a~single supermultiplet which contains the
metric $ds^2=g_{mn} dx^m dx^n$, a~Majorana spinor-valued 1-form gravi\-ti\-no $\psi_m dx^m$ and a~real 4-form f\/ield
strength $F= {1 \over 24} F_{mnpq} dx^m \wedge dx^n \wedge dx^p \wedge dx^q$, with $m,n,p,q=0,1,\dots, 9, \natural$.
The f\/ield~$F$ derives from a~3-form potential~$C$ by $F=dC$ and thus obeys the Bianchi identity $dF=0$.
The f\/ield equations are given by~\cite{Cremmer:1978km}
\begin{gather}
d (\star F)=\half F \wedge F,
\qquad
R_{mn}={1 \over 12} F_{mpqr} F_n{}^{pqr}+{1 \over 144} g_{mn} F_{pqrs} F^{pqrs}
\label{4a1}
\end{gather}
up to terms which vanish as the gravitino f\/ield $\psi_m$ vanishes (and which will not be needed in the sequel).
Here, $R_{mn}$ is the Ricci tensor, and $\star F$ denotes the Poincar\'e dual of~$F$.
The f\/ield equations derive from an action which contains the Einstein-Hilbert term, the standard kinetic term for~$F$
term, and a~Chern--Simons term for~$F$.
We shall not need the action here.

The supersymmetry transformations acting on the gravitino f\/ield are given by~\cite{Cremmer:1978km}
\begin{gather}
\label{4a2}
\delta_\varepsilon \psi_m=D_m \varepsilon+{1 \over 288}\left(\Gamma_m {}^{npqr}-8 \delta_m {}^n \Gamma^{pqr}\right)F_{npqr} \varepsilon
\end{gather}
up to terms which vanish as $\psi_m$ vanishes (and which will not be needed in the sequel).
Here,~$\varepsilon$ is an arbitrary space-time dependent Majorana spinor supersymmetry transformation parameter, $D_m
\varepsilon$ stands for the standard covariant derivative on spinors, the Dirac matrices are def\/ined by the Clif\/ford
algebra relations $\{\Gamma_m, \Gamma_n \}=2 I g_{mn}$, and~$\Gamma$ matrices with several lower indices are
completely anti-symmetrized in those indices.
The supersymmetry transformations of the bosonic f\/ields $(g,F)$ are odd in~$\psi$ and thus vanish as~$\psi$ vanishes,
and they will not be needed here.

\subsection{Supersymmetric solutions}

Classical solutions are usually considered for vanishing Fermi f\/ields, since classically Fermi f\/ields take values in
a~Grassmann algebra and have odd grading.
Thus, we shall set the gravitino f\/ield to zero, $\psi_m=0$.
The f\/ield equations of~\eqref{4a1} now hold exactly.

A classical solution $(g,F)$ is said to be {\it BPS} or {\it supersymmetric} provided there exist a~non-zero
supersymmetry transformations~$\varepsilon$ which preserve the condition $\psi_m=0$, when the bosonic f\/ields
in~\eqref{4a2} are evaluated on the solution in question.
Thus, the central equation in the study of supersymmetric solutions is the so-called BPS-equation
\begin{gather}
\label{4b1}
\left(D_m+{\cal F}_m\right)\varepsilon =0,
\qquad
{\cal F}_m={1 \over 288}\left(\Gamma_m {}^{npqr}-8 \delta_m {}^n \Gamma^{pqr}\right)F_{npqr}.
\end{gather}
The vector space of solutions ${\cal V}_\varepsilon={\cal V}_\varepsilon (g,F) $ for the spinor~$\varepsilon$ depends
upon the values taken by the bosonic f\/ields $(g,F)$.
For the f\/lat Minkowski solution with $F=0$, the dimension of ${\cal V}_\varepsilon$ is maximal and equal to 32,
corresponding to 32 Poincar\'e supersymmetries.
For the M2 or M5 brane solutions of~\eqref{2a2} (along with the corresponding expressions for~$F$ which we shall provide
later), the dimension of ${\cal V}_\varepsilon$ is 16.
For the near-horizon limits of these branes given in~\eqref{2b1}, the supersymmetry of the corresponding space-times
${\rm AdS}_4 \times S^7$ and ${\rm AdS}_7 \times S^4$ is enhanced and the dimension of ${\cal V}_\varepsilon$ is now 32,
and thus equal to the number of fermionic generators of the superalgebras ${\rm OSp}(8|4,{\mathbb R})$ and ${\rm
OSp}(8^*|4)$ respectively.

We shall be interested in obtaining classical solutions with 16 supersymmetries, namely for which $\dim {\cal
V}_\varepsilon=16$.
Such solutions are referred to as half-BPS.

\subsection{Integrability and the BPS equations}

The BPS equations of~\eqref{4b1} consist of 11 equations each of which is a~32-component Majorana spinor.
This system of 352 equations is subject to $1760$ integrability conditions, given~by
\begin{gather}
\label{4c1}
\left({1 \over 4} R_{mnpq} \Gamma^{pq}+D_m {\cal F}_n-D_n {\cal F}_m+[{\cal F}_m, {\cal F}_n]\right)\varepsilon =0,
\end{gather}
where we have used the fact that the commutator of the spin covariant derivates $D_m$ is given by the Riemann tensor,
$[D_m, D_n] \varepsilon={1 \over 4} R_{mnpq} \Gamma^{pq}$.
Note that, as equations in~$\varepsilon$, the integrability conditions~\eqref{4c1} are purely algebraic.
For generic values of the f\/ields~$g$ and~$F$, there will be no solutions to~\eqref{4c1}, since a~generic f\/ield
conf\/iguration is not supersymmetric.

One may investigate the classif\/ication of supergravity conf\/igurations $(g,F)$ which satisfy the integrability conditions
in~\eqref{4c1} {\it for a~given number of supersymmetries} $\dim {\cal V}_\varepsilon$.
This line of attack has proven fruitful, and has given rise to a~number of important theorems.
It is by now well established that requiring maximal supersymmetry, namely $\dim {\cal V}_\varepsilon=32$, leads to
a~small family of solutions, including f\/lat Minkowski space-time, the $AdS\times S$ solutions, and
pp-waves~\cite{FigueroaO'Farrill:2002ft}.
Powerful techniques to analyze the BPS system have been developed in~\cite{Gauntlett:2002fz} based on the exterior
dif\/ferential algebra of forms constructed out of Killing spinors, and in~\cite{Hull:2003mf} based on the structure of
the holonomy group of~\eqref{4b1}.
Increasingly stronger results are being obtained, for example in~\cite{Gran:2010tj}, where it was shown that any
solution with $\dim {\cal V}_\varepsilon \geq 30$ actually has the maximal number of 32 supersymmetries.

An alternative question is for which values of $\dim {\cal V}_\varepsilon$ the BPS integrability conditions guarantee
that a~conf\/iguration $(g,F)$ satisf\/ies the Bianchi identity $dF=0$ and the f\/ield equations~\eqref{4a1}.
Given the results of the preceding paragraph, the answer is af\/f\/irmative for $\dim {\cal V}_\varepsilon \geq 31$.
For the families of solutions with space-time of the form ${\rm AdS}_3 \times S^3 \times S^3 \times \Sigma$ and $\dim
{\cal V}_\varepsilon =16$ considered in the present paper, the answer is also af\/f\/irmative.
Similar results hold in Type IIB solutions with 16 supersymmetries~\cite{D'Hoker:2007xy,D'Hoker:2007xz,D'Hoker:2007fq}.
As far as we know, however, the question is open for general families of solutions with $\dim {\cal V}_\varepsilon=16$.
Finally, for $\dim {\cal V}_\varepsilon < 16$, the BPS equations do not generally imply {\it all} the Bianchi identities
and f\/ield equations, as explicit counter examples are known.

Viewed in terms of integrability conditions which reproduce all the Bianchi and f\/ield equations for $(g,F)$, the BPS
system bears some striking similarities with the Lax systems, or f\/latness conditions, in low-dimensional classical
integrable systems.
The main dif\/ference here is that the dimension is high, namely 11.
The most interesting cases of similarity are when $\dim {\cal V}_\varepsilon$ is large enough for the BPS equations to
imply the Bianchi and f\/ield equations, but small enough to allow for large families of solutions.
It appears that the case $\dim {\cal V}_\varepsilon=16$ satisf\/ies both requirements, as we shall show next.

\section{Solving the Half-BPS equations}\setcounter{equation}{0}\label{sec5}

In this section, we shall show that the BPS equations for the geometry of the half-BPS intersecting branes in the
near-horizon limit may be mapped onto a~classical integrable conformal f\/ield theory in 2 dimensions of the Liouville
sine-Gordon type.

\subsection{The Ansatz for space-time and f\/ields}

The bosonic symmetry algebra ${\rm SO}(2,2) \oplus {\rm SO}(4) \oplus {\rm SO}(4)$ of the half-BPS intersecting brane
conf\/iguration in the near-horizon limit dictates the structure of the space-time manifold of the solution to be of the form
\begin{gather}
\label{5a1}
\left({\rm AdS}_3 \times S_2^3 \times S_3^3\right)\ltimes \Sigma.
\end{gather}
Here, $S_2^3$ and $S_3^3$ are two dif\/ferent 3-spheres,~$\Sigma$ is a~Riemann surface with boundary, and the product
$\ltimes$ is warped in the sense that the radii of the spaces ${\rm AdS}_3$, $ S_2^3$, and $S_3^3$ are all functions of~$\Sigma$.
The action of the isometry algebra ${\rm SO}(2,2) \oplus {\rm SO}(4) \oplus {\rm SO}(4)$ is
on the space ${\rm AdS}_3 \times S_2^3 \times S_3^3$, for every point on~$\Sigma$.

The bosonic f\/ields invariant under ${\rm SO}(2,2) \oplus {\rm SO}(4) \oplus {\rm SO}(4)$ may be parametrized by
\begin{gather}
ds^2=f_1^2 ds^2_{{\rm AdS}_3}+f_2^2 ds^2_{S_2^3}+f_3^2 ds^2_{S_3^3}+ds^2_\Sigma,
\nonumber
\\
F=db_1 \wedge \omega_{{\rm AdS}_3}+db_2 \wedge \omega_{S_2^3}+db_3 \wedge \omega_{S_3^3},
\nonumber
\\
C=b_1 \omega_{{\rm AdS}_3}+b_2 \omega_{S_2^3}+b_3 \omega_{S_3^3}.
\label{5a2}
\end{gather}
Here, $ ds^2_{{\rm AdS}_3}$ is the ${\rm SO}(2,2)$-invariant metric on ${\rm AdS}_3$ with radius 1 and $\omega_{{\rm
AdS}_3}$ is its volume form.
Similarly, $ ds^2_{S^3_a}$ for $a=2,3$ is the ${\rm SO}(4)$-invariant metric on $S_a^3$ with radius 1 and
$\omega_{S^3_a}$ is its volume form.
The functions $f_1$, $f_2$, $f_3$, $b_1$, $b_2$, $b_3$ are real-valued functions of~$\Sigma$, and do not depend on ${\rm AdS}_3
\times S_2^3 \times S_3^3$.
Finally, $ds_\Sigma^2 $ is a~Riemannian metric on~$\Sigma$.

Since the volume forms $\omega_{{\rm AdS}_3}$ and $\omega_{S^3_a}$ are closed, the form~$F$ indeed obeys $F=dC$ which in
turn automatically satisf\/ies the Bianchi identity.

\subsection{Reduced equations}

The BPS equations of~\eqref{4b1} may be restricted to bosonic f\/ields of the form given by the ${\rm SO}(2,2) \oplus {\rm
SO}(4) \oplus {\rm SO}(4)$-invariant Ansatz of~\eqref{5a2}.
The resulting reduced BPS equations are quite involved, but may be reduced to a~dependence on the following data:
\begin{itemize}\itemsep=0in
\item a~real-valued function~$h$ on~$\Sigma$;
\item a~complex-valued function~$G$ on~$\Sigma$;
\item three real constants $c_1$, $c_2$, $c_3$ which satisfy $c_1+c_2+c_3=0$.
\end{itemize}
The functions $f_1$, $f_2$, $f_3$, $b_1$, $b_2$, $b_3$ and the metric $ds^2_\Sigma$ which parametrize the Ansatz may be expressed
in terms of these data with the help of the following composite quantities
\begin{gather*}
\gamma={c_2 \over c_3},
\qquad
W_\pm=| G \pm i |^2+\gamma^{\pm 1} (G \bar G -1).
\end{gather*}
A~lengthy calculation then gives the following expressions for the metric factors
\begin{gather}
f_2^6={h^2 W_-(G \bar G -1) \over c_2^3 c_3^3 W_+^2},
\qquad
f_1^6={h^2 W_+W_-\over c_1^6 (G\bar G -1)^2},
\nonumber
\\
f_3^6={h^2 W_+(G \bar G -1) \over c_2^3 c_3^3 W_-^2},
\qquad
ds_\Sigma^2={|\partial h |^6 W_+W_-(G \bar G-1) \over c_2^3 c_3^3 h^4}.
\label{5b2}
\end{gather}
The product of the metric factors is particularly simple, and given by
\begin{gather}
\label{5b5}
c_1 c_2 c_3 f_1 f_2 f_3=\sigma h,
\end{gather}
where~$\sigma$ may take the values $\pm 1$, and will be further specif\/ied later.
The expressions for the functions $b_1$, $b_2$, $b_3$ will be exhibited in equation~\eqref{5b3} below.

\subsection{Regularity conditions}

The requirements of regularity consist of two parts.
First, we have the condition of reality, positivity, and the absence of singularities for the metric factors $f_1^2$, $f_2^2$, and $f_3^2$
in the interior of~$\Sigma$.
Second, we have regularity conditions on the boundary of~$\Sigma$.
Clearly, these conditions require $c_1$, $c_2$, $c_3$ and~$h$ to be real, as we had already stated earlier, and as we shall
continue to assume in the sequel.

The requirements of regularity in the interior of~$\Sigma$ are as follows.
We must have
\begin{enumerate}\itemsep=0in
\item[1)] positivity of $f_1^6$ which requires $W_+W_-\geq 0$;
\item[2)] positivity of $ds_\Sigma^2$ which requires $\gamma (G \bar G-1) W_+W_-\geq 0$;
\item[3)] positivity of $f_2^6$ and $f_3^6$ which requires $\gamma (G \bar G-1) W_\pm \geq 0$.
\end{enumerate}
A~necessary and suf\/f\/icient condition for all three requirements above to hold true is
\begin{gather}
\label{5c1}
\gamma (G \bar G -1) \geq 0
\end{gather}
as may be readily verif\/ied by inspecting~\eqref{5b2}.

The requirements of regularity at the boundary $\partial \Sigma$ of~$\Sigma$ are more delicate.
We begin by stressing that $\partial \Sigma$ does not correspond to a~boundary of the space-time manifold of the
supergravity solution; rather it corresponds to interior points.
What is special about the points on~$\partial \Sigma$ is that either one or the other three spheres, $S_2^3$ or $S_3^3$
shrinks to zero radius there.
Such points naturally appear when f\/ibering any unit sphere~$S^{n+1}$ over its equal latitude angle~$\theta$ in the
interval $[0,\pi] $.
At each value of~$\theta$, we have a~sphere $S^n$ whose radius varies with~$\theta$ and goes to 0 for $\theta =0,\pi$.
This behavior is manifest from the relation between the unit radius met\-rics~$ds^2_{S^n}$ and $ds^2_{S^{n+1}}$ in this f\/ibration{\samepage
\begin{gather*}
ds^2_{S^{n+1}}=d \theta^2+(\sin \theta)^2 ds^2_{S^n}.
\end{gather*}
From the point of view of the total space $S^{n+1}$ the points $\theta=0, \pi$ are unremarkable.}

In the geometry at hand, the boundary $\partial \Sigma$ is 1-dimensional, and the f\/ibration will enter for $S^4$, $S^7$,
and ${\rm AdS}_7$ (the f\/ibration of ${\rm AdS}_4$ over ${\rm AdS}_3$ has no vanishing points).
In each case, $\partial \Sigma$ corresponds to the vanishing of either $f_2$ or $f_3$, but never of $f_2$ and $f_3$
simultaneously.
Conversely, all points where either $f_2$ or $f_3$ vanishes belong to $\partial \Sigma$.
Applying these considerations to the formulas for the metric factors in~\eqref{5b2} and~\eqref{5b5}, we derive the
following necessary and suf\/f\/icient regularity conditions at the boundary $\partial \Sigma$:
\begin{enumerate}\itemsep=0in
\item[1)] $h =0$ on $\partial \Sigma$ in view of $f_2f_3=0$ there and equation~\eqref{5b5};
\item[2)] $W_+=0$ when $f_3=0$ and $f_2 \not= 0$ from the vanishing of~$h$ on $\partial \Sigma$;
\item[3)] $W_-=0$ when $f_2=0$ and $f_3\not= 0$ from the vanishing of~$h$ on $\partial \Sigma$.
\end{enumerate}
It follows from this that if $h=0$ everywhere on the boundary $\partial \Sigma$, and if we assume the supergravity
solution, and thus~$\Sigma$ to be connected, then the sign of~$h$ must be constant throughout~$\Sigma$.
Without loss of generality, we choose
\begin{gather}
\label{5c2}
h > 0  \qquad \text{in the interior of~$\Sigma$}.
\end{gather}
Finally, we note that the conditions $W_\pm =0$ of points~2) and~3)
above are readily solved under the assumption~\eqref{5c1} with the following result
\begin{gather}
\label{5c3}
W_\pm =0
\qquad
\Leftrightarrow
\qquad
G= \mp i.
\end{gather}

\subsection{Dif\/ferential equations}

The BPS equations~\eqref{4b1} for bosonic f\/ields given by the Ansatz of~\eqref{5a2}, are solved in part by the equations
given in~\eqref{5b2} and~\eqref{5b3}, provided the functions~$h$ and~$G$ satisfy the following dif\/ferential equations
\begin{gather}
\label{5d1}
\partial_w \partial_{\bar w} h =0,
\qquad
h \partial_w G=\half (G+\bar G) \partial_w h
\end{gather}
along with the complex conjugate of the second equation.
Here, $w$, $\bar w$ are local complex coordinates, and the above equations are invariant under conformal reparametrizations
of~$w$.

The second equation in~\eqref{5d1} guarantees the existence (at least locally) of a~real function~$\Phi$, def\/ined by the
dif\/ferential equation
\begin{gather}
\label{5d1a}
\partial_w \Phi=\bar G (\partial_w \ln h).
\end{gather}
The integrability condition between this equation and its complex conjugate is satisf\/ied as soon as the equations
of~\eqref{5d1} are.
In turn,~\eqref{5d1a} and its complex conjugate may be used to eliminate~$G$ and $\bar G$, which gives the following
second order dif\/ferential equation for~$\Phi$
\begin{gather*}
2 \partial_{\bar w} \partial_w \Phi+\partial_{\bar w} \Phi (\partial_w \ln h)-\partial_w \Phi (\partial_{\bar w} \ln h) =0.
\end{gather*}
It must be remembered, of course, that~$G$ must satisfy the inequality~\eqref{5c1} which in terms of~$\Phi$ translates
to a~rather unusual looking inequality, namely $\gamma (|\partial_w \Phi |^2-|\partial_w \ln h|^2) >0$.

\subsection{Flux f\/ield solutions}

The components of the potential~$C$, namely $b_1$, $b_2$, $b_3$ are found to be given as follows
\begin{gather}
b_1=b_1^0+{\nu_1 \over c_1^3}\left({h (G+\bar G) \over G \bar G-1}+\big(2+\gamma+\gamma^{-1}-\big(\gamma-\gamma^{-1}\big)\big) \tilde h\right),
\nonumber
\\
b_2=b_2^0-{\nu_2 \over c_2^2 c_3}\left({h (G+\bar G) \over W_+}-\Phi+\tilde h\right),
\qquad
b_3=b_3^0+{\nu_3 \over c_2 c_3^2}\left({h (G+\bar G) \over W_-}-\Phi+\tilde h\right).
\label{5b3}
\end{gather}
The arbitrary constants parameters $b_1^0$, $b_2^0$, $b_3^0$ account for the residual gauge transformations on the 3-form
f\/ield~$C$ which are allowed within the Ansatz.
The factors $\nu_1$, $\nu_2$, $\nu_3$ may take values~$\pm 1$, but supersymmetry places a~constraint on their product
\begin{gather*}
\sigma=-\nu_1 \nu_2 \nu_3,
\end{gather*}
where $\sigma $ is the sign encountered in equation~\eqref{5b5}.
The real-valued function~$\Phi$ has already been def\/ined in~\eqref{5d1a}, and is determined in terms of~$G$ and~$h$ up
to an additive constant.
The function $\tilde h$ is the harmonic function dual to the harmonic function~$h$ and satisf\/ies
\begin{gather*}
\partial_{\bar w} (h+i \tilde h) =0
\end{gather*}
along with its complex conjugate equation.

The electric f\/ield strength, suitably augmented to a~conserved combination in order to account for the presence of the Chern--Simons
interaction, may be decomposed on the reduced geometry $({\rm AdS}_3 \times S^3 \times S^3) \ltimes \Sigma$, as follows
\begin{gather}
\label{5d3}
\star F+\half C \wedge F=-d \Omega_1 \wedge \hat e^{345678}+d \Omega_2 \wedge \hat e^{678012}+d \Omega_3 \wedge \hat e^{012345}.
\end{gather}
The Bianchi identity $dF=0$ and the f\/ield equation for~$F$ guarantee that the 7-form on each side is a~closed
dif\/ferential form, whence the notations $d\Omega_a$ with $a=1,2,3$, with the understanding that $\Omega_a$ may or may
not be single-valued.
Since only the 6-cycle conjugate to $d\Omega_1$ is compact, we shall focus on its properties, and we f\/ind
\begin{gather*}
\Omega_1={\sigma \nu_1 \over c_2^3 c_3^3}\big(\Omega_1^0+\Omega_1^s+\Lambda-\tilde h \Phi\big).
\end{gather*}
Here, $\Omega_1^0$ is constant, the function $\Lambda $ satisf\/ies the dif\/ferential equation
\begin{gather*}
\partial_w \Lambda=i h \partial_w \Phi-2 i \Phi \partial_w h
\end{gather*}
and the function $\Omega_1^s$ is given by
\begin{gather*}
\Omega_1^s=\sum\limits_{\pm} {h \over 2 W_\pm}\big(\gamma^{\pm 1} h \big(|G|^2-1\big) +(\Phi \pm \tilde h) (G+\bar G)\big).
\end{gather*}
The M2 brane charges of a~solution are obtained by integrating~\eqref{5d3} over compact seven-cycles, which consist of
the warped product of $S_2^3 \times S_3^3$ over a~curve in~$\Sigma$ that is spanned between one point on $\partial
\Sigma$ where $f_3=0$ and another point on $\partial \Sigma $ where $f_3=0$.
These charges give the net numbers of M2 branes ending respectively on M5 and M5$'$ branes.

\section{Map to an integrable system} \label{sec6}

The equations obeyed by~$h$ and~$G$ in the interior of~$\Sigma$ may be summarized as follows For~$h$ we have
\begin{gather}
\label{5e1}
\partial_w \partial_{\bar w} h =0,
\qquad
h > 0,
\end{gather}
while for~$G$ we have
\begin{gather*}
\partial_w G=\half (G+\bar G) \partial_w \ln h,
\qquad
\gamma (G \bar G -1) > 0.
\end{gather*}
The conditions on the boundary of~$\Sigma$ may be summarized as follows
\begin{gather*}
h=0,
\qquad
G=\pm i.
\end{gather*}
These equations are {\it solvable} in the following sense.

For any given~$\Sigma$, one begins by solving for~$h$ and obtaining a~real harmonic function~$h$ that is strictly
positive in the interior of~$\Sigma$ and vanishes on $\partial \Sigma$.
The algorithm for doing so is routine, as the dif\/ferential equation, the positivity condition, and the boundary
condition obey a~superposition principle under addition with positive coef\/f\/icients.
If $h_1$ and $h_2$ are real harmonic, positive inside~$\Sigma$, and vanishing on the boundary $\partial \Sigma$, then so
is $\alpha_1 h_1+\alpha_2 h_2$ for any real positive coef\/f\/icients $\alpha_1$, $\alpha_2$ which are not both zero.

Having solved for~$h$, we now assume that~$h$ is given by one such solution, and we proceed to considering the equations
for~$G$ and $\bar G$, namely the second dif\/ferential equation in~\eqref{5e1} along with its complex conjugate.
For f\/ixed~$h$, these equations are linear in~$G$ provided the superposition is carried out with real coef\/f\/icients.
If $G_1$ and $G_2$ obey the second dif\/ferential equation in~\eqref{5e1}, then so does $a_1 G_1+a_2 G_2$ for any real~$a_1$,~$a_2$.

However, the positivity condition $\gamma (G \bar G -1)>0$ and the boundary condition $G=\pm i$ will not be maintained
by such linear superposition, even if with real coef\/f\/icients.
The key reason is that the f\/irst condition is not linear.
Thus, the linearity of the dif\/ferential equations for~$G$ is obstructed by the non-linearity of the positivity and
boundary conditions, and the full problem is genuinely non-linear.

\subsection{Associated integrable system}

To expose the presence of an integrable system, we parametrize the complex function~$G$ by polar coordinates in terms of
real functions $\psi >0$ and~$\theta$
\begin{gather*}
G=\psi e^{i \theta}.
\end{gather*}
The non-linear constraint then reduces to the linear relation
\begin{gather*}
\gamma (\psi -1) >0  \qquad \text{in the interior of~$\Sigma$}
\end{gather*}
while the boundary conditions are also linear, and given by
\begin{gather*}
\psi =1,
\qquad
\theta=\pm {\pi \over 2}.
\end{gather*}
However, the dif\/ferential equation for~$G$, and its complex conjugate equation, expressed in terms of the
variables~$\psi$ and~$\theta$ are now non-linear, and given by
\begin{gather}
\partial_w \ln \psi+i \partial_w \theta=\big(1+e^{-i \theta}\big)\partial_w \ln h,
\qquad
\partial_{\bar w} \ln \psi-i \partial_{\bar w} \theta=\big(1+e^{+i \theta}\big)\partial_{\bar w} \ln h.
\label{5f4}
\end{gather}
The integrability condition on the system~\eqref{5f4}, viewed as equations for~$\theta$, will involve both $\psi $
and~$\theta$, and will be of no interest here.
When the system~\eqref{5f4} is viewed as equations for~$\psi$, the integrability condition is a~second order equation
for the f\/ield~$\theta$ only, and is given by
\begin{gather}
\label{5f5}
2 \partial_{\bar w} \partial_w \theta+2 \sin \theta (\partial_{\bar w} \partial_w \ln h)+e^{+i \theta} \partial_w
\theta (\partial_{\bar w} \ln h)+e^{-i \theta} \partial_{\bar w} \theta (\partial_w \ln h) =0.
\end{gather}
This equation is integrable in the classical sense.
To see this, one may either interpret~\eqref{5f4} as a~{\it B\"acklund transformation} for the equation~\eqref{5f5}, or
one may expose a~Lax pair associated with~\eqref{5f5}.
That Lax pair does indeed exist, and is given as follows
\begin{gather*}
L_w=\partial_w +A_w,
\qquad
A_w =+i \partial_w \theta -\big(1+e^{-i \theta}\big)\partial_w \ln h,
\\
L_{\bar w}=\partial_{\bar w} +A_{\bar w},
\qquad
A_{\bar w}=-i \partial_{\bar w} \theta -\big(1+e^{+i\theta}\big)\partial_{\bar w} \ln h.
\end{gather*}
Flatness of the connection $A_w$, $A_{\bar w}$ and of the covariant derivatives $L_w$ and $L_{\bar w}$ implies
equation~\eqref{5f5}.
The associated Lax equations
\begin{gather*}
L_w \psi=L_{\bar w} \psi =0
\end{gather*}
coincide with the set of equations~\eqref{5f4} that we started with.
In summary, equation~\eqref{5f5} has an associated Lax pair, and is integrable in the classical sense.

Furthermore, equation~\eqref{5f5} is invariant under conformal reparametrizations of the local complex coordinates~$w$
and $\bar w$.
Using this invariance, one may choose local complex coordi\-na\-tes~$w$,~$\bar w$ such that $h=\Im(w)$, so that
equation~\eqref{5f5} becomes
\begin{gather*}
2 \partial_{\bar w} \partial_w \theta+{2 \over (w-\bar w)^2} \sin \theta-{1 \over w-\bar w} e^{+i \theta}
\partial_w \theta+{1 \over w -\bar w} e^{-i \theta} \partial_{\bar w} \theta =0.
\end{gather*}
This equation now depends upon a~single real f\/ield~$\theta$ and is clearly related to the sine-Gordon and Liouville
equations~\cite{D'Hoker:1982er}, specif\/ically the Liouville equation on the upper half plane with the Poincar\'e
constant negative curvature metric
\begin{gather*}
ds^2_\Sigma={|dw|^2 \over \Im (w)^2}
\end{gather*}
as discussed, for example, in~\cite{D'Hoker:1983is}.

\section[Role of the superalgebra $D(2,1;\gamma) \oplus D(2,1;\gamma)$]{Role of the superalgebra $\boldsymbol{D(2,1;\gamma) \oplus D(2,1;\gamma)}$}

Earlier in this paper, we have stated the expectation that the supersymmetries of the half-BPS solution of intersecting
M2, M5, and M5$'$ branes in the near-horizon limit will generate the Lie superalgebra $D(2,1;\gamma) \oplus
D(2,1;\gamma)$.
In this section, we shall review the structure of the Lie superalgebra $D(2,1;\gamma)$, list some of its properties, and
show that it is indeed realized by the solutions obtained above.

\subsection[The Lie superalgebra $D(2,1;\gamma)$]{The Lie superalgebra $\boldsymbol{D(2,1;\gamma)}$}

The complex Lie superalgebra $D(2,1;\gamma)$ is the only f\/inite-dimensional simple Lie superalgebra that depends on
a~continuous parameter, namely the complex parameter~$\gamma$.
The maximal bosonic subalgebra of $D(2,1;\gamma)$ is ${\rm SL}(2,{\mathbb C}) \oplus {\rm SL}(2,{\mathbb C}) \oplus {\rm SL}(2,{\mathbb C})$.
The smallest classical Lie superalgebra which contains $D(2,1;\gamma)$ for all values of~$\gamma$ is ${\rm OSp}(9|8)$.
An equivalent way of representing~$\gamma$ is by three complex numbers $c_1$, $c_2$, $c_3$ modulo an overall complex
rescaling, which satisfy $c_1+c_2+c_3=0$ and $\gamma=c_2/c_3$.
The six permutations $\sigma \in \mathfrak{S}_3$ of the numbers $c_1$, $c_2$, $c_3$
induce permutations $\sigma (\gamma)=c_{\sigma (2)} / c_{\sigma (3)}$ under which the complex algebra is invariant
\begin{gather*}
D(2,1;\sigma (\gamma))=D(2,1;\gamma).
\end{gather*}
The complex Lie superalgebra $D(2,1;\gamma)$ has three inequivalent real forms which are denoted $D(2,1;\gamma,p)$ for
$p=0,1,2$, and which have~$\gamma$ real and maximal bosonic subalgebra ${\rm SO}(2,1) \oplus {\rm SO}(4-p,p)$.
The real form of interest here is $D(2,1;\gamma,0)$; its maximal bosonic subalgebra is isomorphic to ${\rm SO}(2,1)
\oplus {\rm SO}(3) \oplus {\rm SO}(3)$.
The automorphism group $\mathfrak{S}_3$ is reduced to the subgroup~$\mathfrak{S}_2$ which permutes the two ${\rm SO}(3)$
algebras, and acts by $\sigma (\gamma)=\gamma^{-1}$.

The generators of the maximal bosonic subalgebra\footnote{The labels on the simple factors are introduced in analogy
with the notation of the corresponding factor spaces in~\eqref{5a1} and~\eqref{5a2}.} ${\rm SO}(2,1)_1 \oplus {\rm
SO}(3)_2 \oplus {\rm SO}(3)_3$ of the real form $D(2,1;\gamma,0)$ will be denoted by $T_i^{(a)}$, where the index
$a=1,2,3$ refers to the simple components of the algebra and the index $i=1,2,3$ labels the three generators
corresponding to component~$a$.
For example, $T^{(1)}_i$ are the three generators of ${\rm SO}(2,1)_1$.
The bosonic structure relations are given as follows
\begin{gather*}
\big[T^{(a)}_i, T^{(b)}_j\big]=i \delta^{ab} \varepsilon_{ijk} \eta_{(a)}^{k\ell} T^{(a)}_\ell,
\end{gather*}
where $\eta_{(2)}=\eta_{(3)}={\rm diag} (+++)$ are the invariant metrics of ${\rm SO}(3)_2$ and ${\rm SO}(3)_3$ and
$\eta_{(1)}={\rm diag} (-, +, +)$ is the invariant metric of ${\rm SO}(2,1)_1$.

The fermionic generators of $D(2,1;\gamma,0)$ transform under the 2-dimensional irreducible representation of each one
of the bosonic simple subalgebras.
We shall denote these generators by~$F$ with components $F_{\alpha_1, \alpha_2, \alpha_3}$ where $\alpha_a $ are
2-dimensional spinor indices.
This characterization uniquely determines the commutation relations of $T^{(a)}$ with~$F$.
The remaining structure relations are given by the anti-commutators of~$F$ which take the form
\begin{gather*}
\{F_{\alpha_1, \alpha_2, \alpha_3}, F_{\beta_1, \beta_2, \beta_3} \}
= c_1 (C\sigma^i)_{\alpha_1 \beta_1}C_{\alpha_2 \beta_2} C_{\alpha_3 \beta_3} T_i^{(1)}
\\
\hphantom{\{F_{\alpha_1, \alpha_2, \alpha_3}, F_{\beta_1, \beta_2, \beta_3} \}=}{}
  +c_2 C_{\alpha_1 \beta_1} (C\sigma^i)_{\alpha_2 \beta_2} C_{\alpha_3 \beta_3} T_i^{(2)}
  +c_3 C_{\alpha_1 \beta_1} C_{\alpha_2 \beta_2} (C\sigma^i)_{\alpha_3 \beta_3} T_i^{(3)}.
\end{gather*}
Here, $\sigma^i$ are the Pauli matrices, $C=i \sigma^2$, $\gamma=c_2/c_3$, and $c_1+c_2+c_3=0$.
For the complex Lie superalgebra $D(2,1;\gamma)$, the parameters $c_1$, $c_2$, $c_3$ are complex while for the real forms
these parameters are real.
For the real form $D(2,1;\gamma,0)$, the automorphism $\gamma \to \gamma^{-1}$ amounts to interchanging the generators
with labels $a=2$ and $a=3$.

\subsection[Invariance of the solutions under $D(2,1;\gamma,0) \oplus D(2,1;\gamma,0)$]{Invariance
of the solutions under $\boldsymbol{D(2,1;\gamma,0) \oplus D(2,1;\gamma,0)}$}

In this subsection, we shall summarize the arguments of Appendix D of~\cite{D'Hoker:2008ix} in which a~proof is given of
the invariance under $D(2,1;\gamma,0) \oplus D(2,1;\gamma,0)$ of the half-BPS supergravity solutions.
It is manifest that all solutions have 9 Killing vectors $v^m$ which correspond to the 9 generators of ${\rm SO}(2,1)_1
\oplus {\rm SO}(3)_2 \oplus {\rm SO}(3)_3$, and which satisfy
\begin{gather}
\label{6a2}
\nabla_m v_n+\nabla_n v_m=0.
\end{gather}
By construction, the half-BPS solutions also have 16 supersymmetries, or Killing spinors~$\varepsilon$.
Nine Killing vectors and sixteen Killing are precisely the correct numbers of bosonic and fermionic generators needed
for $D(2,1;\gamma,0) \oplus D(2,1;\gamma,0)$.
To prove that these Killing vectors and Killing spinors together generate the algebra $D(2,1;\gamma,0) \oplus
D(2,1;\gamma,0)$, we must ensure that the structure relations of $D(2,1;\gamma,0) \oplus D(2,1;\gamma,0)$ are satisf\/ied.
This is manifest for the commutation relations of two bosonic generators, and of one bosonic and one fermionic
generator.
Thus, it remains to show that the composition of two fermionic generators gives the bosonic generators {\it with the
correct parameters $c_1$, $c_2$, $c_3$}.

To obtain the composition law for two Killing spinors~$\varepsilon$ and $\varepsilon'$, one proceeds as follows.
Let~$\varepsilon$ and $\varepsilon'$ satisfy the BPS equations~\eqref{4b1}, so that they are Killing spinors.
One proves, using the same BPS equations, that one has
\begin{gather*}
\nabla_m\left(\bar \varepsilon \Gamma_n \varepsilon'\right)={1 \over 3}\left(\bar \varepsilon \Gamma^{pq}\varepsilon'\right)F_{mnpq}.
\end{gather*}
By symmetrizing both sides in~$m$ and~$n$, and using the anti-symmetry of~$F$ in these indices, one readily f\/inds that
the combination
\begin{gather*}
v_m=\bar \varepsilon \Gamma_m \varepsilon'
\end{gather*}
satisf\/ies the Killing vector equation~\eqref{6a2}.

An overall rescaling of $\varepsilon$, $\varepsilon'$ and $v_m$ is immaterial in establishing the structure relations of
$D(2,1;\gamma,0) \oplus D(2,1;\gamma,0)$, but the relative normalizations of the various generators are crucial to
extract the correct ratio~$\gamma$.
Proper normalization may be enforced as follows.
We begin by introducing an invariant basis of 2-component Killing spinors $\chi_1^{\eta_1}$, $\chi_2^{\eta_2}$, and
$\chi_3^{\eta_3}$ respectively of ${\rm SO}(2,1)_1$, ${\rm SO}(3)_2$, and $ {\rm SO}(3)_3$ for $\eta_a=\pm 1$ and
$a=1,2,3$.
The Killing spinors are normalized as follows
\begin{gather*}
-i \bar \chi_1^{\eta} \chi_1^{\eta'}=(\chi_2^\eta)^\dagger \chi_2^{\eta'}=(\chi_3^\eta)^\dagger \chi_3^{\eta'}=\delta^{\eta \eta'}.
\end{gather*}
We decompose the 32-component spinor~$\varepsilon$ on the tensor product of these basis Killing spinors
\begin{gather*}
\varepsilon=\chi_1^{\eta_1} \otimes \chi_2^{\eta_2} \otimes \chi_3^{\eta_3} \otimes \zeta_{\eta_1, \eta_2, \eta_3},
\end{gather*}
where, for each assignment of $\eta_a$, the spinor~$\zeta$ has 4 components.
To evaluate the metric factors and the Killing vectors in terms of~$\zeta$, we introduce an adapted basis of Dirac matrices,
in frame basis, with indices $\mathfrak{a}_1=0,1,2$, $\mathfrak{a}_2=3,4,5$, $\mathfrak{a}_3 =6,7,8$, and $\mathfrak{a}=9, \natural $,
given~by
\begin{gather*}
\Gamma^{\mathfrak{a}_1}=\gamma^{\mathfrak{a}_1} \otimes I_2 \otimes I_2 \otimes \sigma^1 \otimes \sigma^3,
\qquad
\sigma^1=\gamma^1=\gamma^4= \gamma^7=\gamma^9,
\\
\Gamma^{\mathfrak{a}_2}=I_2 \otimes \gamma^{\mathfrak{a}_2} \otimes I_2 \otimes \sigma^2 \otimes \sigma^3,
\qquad
\sigma^2=-i \gamma^0=\gamma^3= \gamma^6=\gamma^\natural,
\\
\Gamma^{\mathfrak{a}_3}=I_2 \otimes I_2 \otimes \gamma^{\mathfrak{a}_3} \otimes \sigma^3 \otimes \sigma^3,
\qquad
\sigma^3=\gamma^2 =\gamma^5=\gamma^8,
\\
\Gamma^{\mathfrak{a}}=I_2 \otimes I_2 \otimes I_2 \otimes I_2 \otimes \gamma^\mathfrak{a}.
\end{gather*}
The relations between the Killing spinors and vectors may be expressed in this manner as well.
We have $\bar \varepsilon \Gamma^{\mathfrak{a}} \varepsilon'=0$, as well as, for $\eta =\pm $
\begin{gather*}
\bar \varepsilon \Gamma^{\mathfrak{a}_1} \varepsilon'=+2 c_1 f_1 \big(\bar \chi_1^{\eta_1} \gamma^{\mathfrak{a}_1}\chi_1^{\eta_1}\big),
\qquad
\bar \varepsilon \Gamma^{\mathfrak{a}_2} \varepsilon'=-2 i c_2 f_2 \big((\chi_2^{\eta_2})^\dagger\gamma^{\mathfrak{a}_2} \chi_2^{\eta_2}\big),
\\
\bar \varepsilon \Gamma^{\mathfrak{a}_3} \varepsilon'=-2 i c_3 f_3 \big((\chi_3^{\eta_3})^\dagger\gamma^{\mathfrak{a}_3} \chi_3^{\eta_3}\big).
\end{gather*}
The Killing vectors $v_a^{\eta_a}$ are related to the normalized Killing vectors $\hat v^{\mathfrak{a}_a}$ and to the
normalized Killing spinors as follows
\begin{gather*}
(v_1^\eta)^{\mathfrak{a}_1}=f_1 (\hat v_1^\eta)^{\mathfrak{a}_1}=f_1 \bar \chi_1^{\eta_1}\gamma^{\mathfrak{a}_1} \chi_1^{\eta_1},
\qquad
(v_2^\eta)^{\mathfrak{a}_2}=f_2 (\hat v_2^\eta)^{\mathfrak{a}_2}=f_2 (\chi_2^{\eta_2})^\dagger\gamma^{\mathfrak{a}_2} \chi_2^{\eta_2},
\\
(v_3^\eta)^{\mathfrak{a}_3}=f_3 (\hat v_3^\eta)^{\mathfrak{a}_3}=f_3 (\chi_3^{\eta_3})^\dagger\gamma^{\mathfrak{a}_1} \chi_3^{\eta_3}
\end{gather*}
so that the physical Killing vectors $v^A$ are gives as follows
\begin{gather*}
\bar \varepsilon \Gamma^{\mathfrak{a}_1} \varepsilon'=2 c_1 (v_1^{\eta_1})^{\mathfrak{a}_1},
\qquad
\bar \varepsilon \Gamma^{\mathfrak{a}_2} \varepsilon'=2 c_2 (v_2^{\eta_2})^{\mathfrak{a}_2},
\qquad
\bar \varepsilon \Gamma^{\mathfrak{a}_3} \varepsilon'=2 c_3 (v_3^{\eta_3})^{\mathfrak{a}_3}.
\end{gather*}
The parameters $c_1$, $c_2$, and $c_3$ now naturally emerge in the structure relations, and are seen to reproduce those of
the algebra $D(2,1;\gamma,0)_+\oplus D(2,1;\gamma,0)_-$ for the indices $\eta=\pm $.
This concludes the proof of the invariance of the solutions under this Lie superalgebra.

\section{Families and moduli spaces of exact solutions}

In this section, we shall give a~brief overview of the solutions that have been derived with the help of the approach
developed in the preceding sections of this paper, and refer the reader to Sections~7 and~8 of~\cite{Bachas:2013vza} for
details and derivations.
We begin with four basic results.
\begin{enumerate}\itemsep=0.0in
\item All solutions arise as families in the parameter $\gamma \in [-1,+1]$, which governs the dependence of the
supergravity f\/ields on the~$\gamma$-independent data $(h,G)$.
\item Across the value $\gamma =0$ the number of $M5$ branes tends to zero, which leads to the decompactif\/ication of the
directions $3456$ in Table~\ref{table2}, which in turn corresponds to sending the radius of the sphere $S^3_2$ to inf\/inity.
(A mirror image decompactif\/ication takes place at $\gamma =\infty$, where the number of M5$'$ branes tends to zero and the
radius of $S_3^3$ diverges.) Across the value $\gamma =-1$ all three components decompactify and the spheres are permuted into one another.
\item Solution for which $\Sigma $ is compact without boundary have positive~$\gamma$, constant $h$, $G$, constant $f_1$, $f_2$, $f_3$,
with $f_3^2=\gamma f_2^2$ and $f_1^2= \gamma f_2^2/(1+\gamma)$.
In this case, the expression for the metric $ds_\Sigma^2$ needs to be def\/ined with some extra care as the naive form
in~\eqref{5b2} would vanish for constant~$h$.
\item A~regular solution with $\gamma <0$ cannot have more than one asymptotic $AdS$ region, and therefore cannot be
dual to an interface conformal f\/ield theory.
\end{enumerate}

An important implication of point~3)
above is the uniqueness of solutions dual to two-dimensional conformal f\/ield theories, namely they must be of the form
${\rm AdS}_3 \times S^3 \times S^3 \times T^2$, where $T^2$ is a~f\/lat two-torus, and the product is direct.
This is the near-horizon geometry of M2 branes suspended between M5 branes, in the limit where the M5 branes have been
smeared~\cite{Boonstra:1998yu}.
The implication of point~3)
is that the infrared dynamics on the M2 branes always restores the translation symmetry which would otherwise have been
broken by te localized M5 branes.
This dynamical behavior should be contrasted with the analogous situation of D3 branes suspended between NS5 branes and
D5 branes, where the 3-dimensional f\/ield theory on the suspended D3 branes has many strongly-coupled infrared f\/ixed
points~\cite{Gaiotto:2008ak, Gaiotto:2008sa,Gaiotto:2008sd}, which are in one-to-one correspondence with a~rich set of
half-BPS solutions of the Type IIB supergravity equations~\cite{Aharony:2011yc, Assel:2011xz,Assel:2012cj}.

Under the regularity conditions spelled out in section~\ref{sec5}, for the case where~$\Sigma$ has a~boundary and away
from the values $\gamma =0,-1$, the solutions of the dif\/ferential equations~\eqref{5d1a} and the boundary
conditions~\eqref{5c2} and~\eqref{5c3} produce regular, fully back-reacted supergravity near-horizon limits of half-BPS
intersecting M2, M5, and M5$'$ branes.

\subsection[Global solutions with $\gamma >0$]{Global solutions with $\boldsymbol{\gamma >0}$}

The maximally supersymmetric solution ${\rm AdS}_4 \times S^7$ corresponds to $\gamma=1$.
It admits a~deformation by one real continuous parameter~$\lambda$ to a~Janus solution, discovered in~\cite{janus}.
Mapping~$\Sigma$ to the upper half complex plane with the real axis as boundary, and using global complex coordina\-tes~$w$,~$\bar w$ on~$\Sigma$, the functions $(h,G)$ are given as follows
\begin{gather*}
h=i\left({1 \over w}-w\right)+{\rm c.c.},
\qquad
G=i {|w|+|w|^{-1}+\lambda (w-\bar w) |w|^{-1} \over \bar w+\bar w^{-1}}.
\end{gather*}
Here,~$\lambda$ can take any real value.
The undeformed ${\rm AdS}_4 \times S^7$ solution corresponds to $\lambda =0$.
The corresponding metric functions $f_1$, $f_2$, $f_3$, $ds^2_\Sigma$ may be obtained for any $\gamma >0$ from the equations
of~\eqref{5b2}, and the f\/lux potentials $b_1$, $b_2$, $b_3$ may be obtained from~\eqref{5b3}.
The solutions describe a~superconformal domain wall of the dual gauge theory on the M2 brane.
This two-parameter deformation of ${\rm AdS}_4 \times S^7$ was discovered by independent means as a~solution of gauged
4-dimensional supergravity in~\cite{Bobev:2013yra}, with whom we f\/ind precise agreement.

\subsubsection{String and semi-inf\/inite M2 branes}

Besides the strictly regular Janus solutions of the preceding paragraph, there are also solution with a~mild singularity
which have non-vanishing M5 charge and are analogous to the highly curved NS5 brane and D5 brane regions
of~\cite{Aharony:2011yc, Assel:2011xz,Assel:2012cj}.
Mapping~$\Sigma$ again to the upper half plane with complex coordinates $w$, $\bar w$, the functions $(h,G)$ are given by
\begin{gather*}
h=-i (w-\bar w),
\qquad
G=\pm G_0 \pm \sum\limits_{n=1}^{N+1} {\zeta_n \Im (w) \over (\bar w-x_n) |w-x_n|}.
\end{gather*}
Here,~$N$ is a~positive integer, and $x_n$ and $\zeta_n$ are arbitrary real moduli of the solutions.
The function $G_0$ may take the values of either $G_0=i$, in which case the conformal boundary of the solution is that
of a~deformed ${\rm AdS}_7 \times S^4$ space-time, or $G_0$ may take the value $G_0=i w/|w|$, in which case the
conformal boundary of the solution is that of a~deformed ${\rm AdS}_4/{\mathbb Z}_2 \times S^7$ space-time.
These solutions describe self-dual strings respectively on the world-volume of M5 branes, or semi-inf\/inite M2 branes.

\subsection[Global solutions with $\gamma <0$]{Global solutions with $\boldsymbol{\gamma <0}$}

The maximally supersymmetric solution ${\rm AdS}_7 \times S^4$ has $\gamma=-\half$, and corresponds to the
near-horizon limit of a~stack of parallel M5 branes.
Mapping $\Sigma $ again to the upper half plane with complex coordinates $w$, $\bar w$, the functions $(h,G)$ for this
solution are given by
\begin{gather}
\label{8b1}
h=-i(w -\bar w),
\qquad
G=i\left(-1+{w +1\over |w+1|}+{w-1 \over |w-1}\right).
\end{gather}
The ef\/fects of taking~$\gamma$ away from the point $\gamma =-\half$ include a~deformation of the metric away from the
maximally symmetric metric, to one that has only the reduced symmetry ${\rm SO}(2,2) \oplus {\rm SO}(4) \oplus {\rm SO}(4)$,
as well as turning on the f\/lux f\/ields associated with M5$'$ and M2 branes, but without generating either a~net
magnetic or electric charge.

\subsubsection{Self-dual strings and Young tableaux}

The solution of~\eqref{8b1} was generalized in~\cite{D'Hoker:2008qm} by the addition of an arbitrary number $2N$ of cap
singularities in~$G$ and, for~$\Sigma$ the upper half plane with coordinates $w$, $\bar w$, is given by
\begin{gather}
\label{8b2}
iG=1+\sum\limits_{n=1}^ {2N+2} (-)^n {w -\xi_n \over | w-\xi_n|}
\end{gather}
for a~set of $2N+2$ real points $\xi_n \in {\mathbb R}$ subject to the ordering $\xi_1 < \xi_2 < \dots < \xi_{2N+2}$.
The parameter~$\gamma$ is allowed to take any negative value.
The solutions are asymptotic to a~single copy of ${\rm AdS}_7 \times S^4$, and correspond to M2 branes ending on, or
intersecting with, M5 and M5$'$ branes.
The holographic duals of these solutions correspond to conformal defects in the 6-dimensional (2,0) theory, and more
concretely arise from the insertion of surface operators in that theory.

More concretely, the space-time manifold of the solutions has $2N+1$ independent non-contractible 4-cycles which are
topologically 4-spheres.
With the sign convention adopted for~$G$ in~\eqref{8b2}, there are $N+1$ cycles supporting M5 brane charges, and~$N$
cycles supporting M5$'$ brane charges.
On~$\Sigma$, the end points of the corresponding curves are the boundary $\partial \Sigma$.
A~basis may be chosen in terms of curves which join the boundary to the left of $\xi_n$ to a~point on the boundary to
the right of $\xi_{n+1}$.
When $n=2a$ is even we obtain a~M5$'$ charge $\mathfrak{M}^{(2)}_a$, while when $n=2b-1$ is odd, we obtain a~M5 charge
$\mathfrak{M}^{(3)}_b$, given by
\begin{gather*}
\mathfrak{M}_a^{(2)}={4 \nu_2 \gamma \over c_2^3}\left(\xi_{2a+1}-\xi_{2a}\right), \qquad a=1,\dots, N,
\\
\mathfrak{M}_b^{(3)}={4 \nu_3 \gamma \over c_3^3}\left(\xi_{2b}-\xi_{2b-1}\right), \qquad b=1,\dots, N+1.
\end{gather*}
These charges determine a~Young tableau, and a~corresponding irreducible representation of ${\rm SU}(N)$.
This Young tableau was conjectured in~\cite{D'Hoker:2008qm} to describe a~surface operator in the corresponding
representation of ${\rm SU}(N)$, in analogy with the behavior of holographic Wilson lines in 4-dimensional ${\cal N}=4$
super Yang--Mills theory~\cite{Gomis:2006sb,Okuda:2008px, Yamaguchi:2006te}.

\section{Summary and open problems}

We have produced an explicit map between the BPS equations for half-BPS solutions of intersecting M2, M5, and M5$'$ branes
in the near-horizon limit, and a~2-dimensional f\/ield theory, which we have shown to be integrable, and to
possess a~Lax pair.
We have constructed a~large family of solutions to the integrable system and to the supergravity BPS equations, and
given their physical interpretation, as far as is presently known.

However, a~full study of the integrable system has not been carried out yet, and is presently the biggest obstacle to
a~complete classif\/ication of all half-BPS solutions of intersection M2, M5, and M5$'$ branes in the near-horizon limit.
We hope that, in this brief article, the mechanics of this integrable system, and the motivation for its further study,
have been clearly exposed, so that the purpose of a~systematic investigation is now clearly def\/ined.

We conclude by stressing that the analogous problem in 10-dimensional Type IIB on ${\rm AdS}_4 \times S^2 \times S^2
\ltimes \Sigma$ or on ${\rm AdS}_2 \times S^4 \times S^2 \ltimes \Sigma$ and 6-dimensional Type 4b supergravities on
${\rm AdS}_2 \times S^2 \ltimes \Sigma$ are, from some points of view, better understood.
This is due, in large part, to the fact that the fermion contents of those supergravities is chiral so that the partial
dif\/ferential equations for various reduced metric and f\/lux components become genuine Cauchy--Riemann equations which may
be solved in terms of holomorphic or harmonic functions.
Indeed, all the supergravities mentioned earlier may be solved in terms of harmonic functions subject to certain
constraints~\cite{Chiodaroli:2011nr, D'Hoker:2007xy,D'Hoker:2007fq}.

\subsection*{Acknowledgments}

It is a~pleasure to thank Constantin Bachas, John Estes, Michael Gutperle, Drya Krym, and Paul Sorba for fruitful
collaborations on the projects reviewed in this paper.
I gratefully acknowledge the warm hospitality and the f\/inancial support of the Laboratoire de Physique Th\'eorique at
the Ecole Normale Sup\'erieure, where part of this work was carried out.
Finally, I would like to acknowledge the organizers, Decio Levi, Willard Miller, Yvan Saint-Aubin, and Pavel Winternitz
for inviting me to the enjoyable conference and celebration in honor of Luc Vinet at the Centre de Recherches
Math\'ematiques.

\pdfbookmark[1]{References}{ref}
\LastPageEnding

\end{document}